\documentclass[prc,twocolumn,groupedaddress,showpacs]{revtex4}
\usepackage{amsmath}
\usepackage{amssymb}
\usepackage{graphicx}

\begin{document}

\preprint{APS/123-QED}

\title{Experimental proposal for accurate determination of the phase relaxation time and testing a formation of thermalized non-equilibrated matter in highly excited quantum many-body systems}

\author{M. Bienert$^1$}

\author{J. Flores$^1$}%

\author{S. Yu. Kun$^{1,2,3}$}

\affiliation{$^1$Centro de Ciencias F\'{i}sicas, Universidad Nacional
  Aut\'{o}noma de M\'{e}xico,Cuernavaca, Morelos, Mexico } 
\affiliation{ $^2$Centre for Nonlinear Physics, RSPhysSE, ANU, Canberra ACT 0200, Australia} \affiliation{ $^3$Department of Theoretical Physics, RSPhysSE, ANU, Canberra ACT 0200, Australia}

\date{\today}

\begin{abstract}
We estimate how accurate the phase relaxation time of quantum many-body systems can be determined from data on forward peaking of evaporating protons from a compound nucleus.
The angular range and accuracy of the data needed for a reliable determination of the phase relaxation time are evaluated. The general method is applied to analyze the inelastic scattering of 18 MeV protons from Pt for which previously measured double differential cross sections for two angles in the evaporating domain of the spectra show a strong forward peaking. A new experiment for an improved determination of the phase relaxation time is proposed. The experiment is also highly desirable for an accurate test of a formation of thermalized non-equilibrated matter in quantum many-body systems.
\end{abstract}

\pacs{24.10.Cn; 24.60.Dr; 25.40.Ep}

\maketitle

For some years one of us has emphasized that phase relaxation in a many-body system can be considerably longer than
 energy relaxation among independent particle states \cite{kun:93,kun:94,kun:97b}. Maybe the easiest experimental access to the
 problem can be found in low energy (10-80 MeV) nucleon-nucleus scattering processes, where some of the oldest data have been
 available for half a century \cite{gugelot}.
It turns out \cite{flores}
that the phase relaxation time, $\tau_{\rm ph}=\hbar /\beta$ with $\beta$ being the phase relaxation width \cite{kun:93,kun:94,kun:97b}, 
is considerably longer than the energy relaxation time,  
$\tau_{\rm erg}=\hbar/\Gamma_{\rm spr}$,
 obtained from standard estimates of the spreading width $\Gamma_{\rm spr}$ of independent particle states.

The significance of this problem results from the fact that an existence of long--living phase relations is of fundamental 
importance in the study of relaxation phenomena in nuclear, atomic, molecular and mesoscopic many--body systems, and for many--qubit
 quantum computation. In particular, if a phase relaxation time, which characterizes the lifetime
 of the ``phase memory'', is longer than the energy relaxation time, this effect could extend
 the time for quantum computing \cite{flores,sigma} beyond the quantum--chaos border \cite{georgeot}.

 The experimental result \cite{gugelot}
 showing that the intensity of the compound nucleus evaporation for $\theta=60^\circ $ exceeds that for $\theta=150^\circ$ by a
 factor of $\simeq$6 indicates that $\beta\sim \Gamma^\uparrow = \hbar/\tau^\uparrow $, where 
 $\Gamma^\uparrow $ is the compound nucleus decay width and  $\tau^\uparrow $ is the compound nucleus life-time.
However, since the evaporation spectra \cite{gugelot} were measured for two angles only, 
these data allow only a rough estimate, $\beta /\Gamma_{\rm spr}\sim 10^{-5}$ \cite{flores}.
The purpose of this note is to determine whether more detailed experiments of the same type could improve this estimate.

Our analysis for the improvement of the experimental determination of the phase relaxation time from nuclear evaporation data will be quite general and can readily be applied to any low energy nucleon-nucleus scattering showing forward peaking in the evaporation domain of the spectra.
 Yet, to be specific, we shall show numbers and graphs for inelastic scattering of 18 MeV protons from Pt, {\it i.e.} for improvement of one of the oldest experiments \cite{gugelot}. The questions we shall answer are: What is the best realistic accuracy of determination of the phase relaxation time from the data on forward peaking of evaporating protons from the compound nucleus? What is the angular range and accuracy of data needed for a reliable determination of $\tau_{\rm ph}$?
An advantage of the analysis is that it relies on relative values of the double differential cross sections which are usually determined experimentally with better accuracy than absolute cross sections.

The proposed experiment is also desirable for an accurate test of the formation of thermalized non-equilibrated matter in compound processes. A manifestation of such a new form of matter, introduced in Refs. \cite{kun:94,kun:97b}, would be equal slopes, {\it i.e.} nuclear ``temperatures''  \cite{blatt}, of the properly scaled \cite{gugelot} proton evaporation spectra for forward and backward angles. 
 Due to the insufficient statistics the data \cite{gugelot} indicate only approximate equality of the slopes for the forward and backward angles with about 20 percent uncertainty.
 We again point out that the proposed test only requires relative values of the proton emission intensities.

The evolution of a nuclear reaction is usually considered to proceed via a series of two-body nucleon-nucleon collisions, which successively form states of increasing complexity. On each stage of the reaction a distinction is made between continuum states and quasibound states. Emissions from the continuum states is related to
  multistep direct reactions \cite{feshbach,tamura,nishioka88}, and decay of the quasibound states originates multistep compound processes \cite{feshbach,nishioka86}. The compound nucleus is formed at the last stage corresponding to the most complex configuration of the chain of quasibound states. The multistep direct reactions originate from the decay of the simplest configurations of the chain resulting in forward-peaked angular distributions. 
In contrast, the multistep pre--compound and compound reactions are conventionally assumed to give rise to angular distributions symmetric about 90 degrees.

We use the exciton model \cite{blann} to evaluate the relative contributions of multistep direct, multistep pre--compound and compound nucleus processes for the p+Pt ($E_p=18$ MeV) inelastic scattering for the proton outgoing energy of 7 and 6 MeV. Fitting the entire energy range for forward angles we found \cite{ourpaper} that, for the proton outgoing energy of 7 MeV, the compound nucleus cross section constitutes 90$\%$, while multistep direct and multistep pre--compound are about 5$\%$ each. 
For the proton outgoing energy of 6 MeV, the compound nucleus cross section is about 98$\%$.
Therefore we observe that even though the low energy 6-7 MeV outgoing proton spectrum is overwhelmingly dominated by compound reactions,
the angular distribution is strongly forward peaked. Clearly a description of the decay of such thermalized but non-equilibrated matter
requires a major modification of conventional theory of compound nucleus (see {\it e.g.} Ref.\cite{blatt}) originally formulated by Bohr, Bethe, Weisskopf,
Wigner, Dyson and others. The basic assumption of the conventional theory is that thermalization of the compound
nucleus guarantees a complete loss of memory of initial phase relations.
A modification of this conventional picture of the compound nucleus was proposed by one of us in Refs. \cite{kun:94,kun:97b}. 
The key element
in the description of asymmetry of angular distributions around 90$^\circ$ c.m. for evaporating particles is total spin 
off-diagonal correlation
between compound nucleus partial width amplitudes. Such a correlation is neglected in a conventional picture of compound nucleus.
Following \cite{kun:94,kun:97b} we have
\begin{alignat}{1}
&\frac{\overline{\gamma_{\mu_1}^{J_1\pi_1a_1}
\gamma_{\mu_1}^{J_1\pi_1b_1}\gamma_{\mu_2}^{J_2\pi_2a_2}
\gamma_{\mu_2}^{J_2\pi_2b_2}}{~}}{
[\overline{(\gamma_{\mu_1}^{J_1\pi_1a_1})^2}{~~}
\overline{(\gamma_{\mu_1}^{J_1\pi_1b_1})^2}{~~}\overline{(\gamma_{\mu_2}^{J_2\pi_2a_2})^2}{~~}
\overline{(\gamma_{\mu_2}^{J_2\pi_2b_2})^2}]^{1/2}}\nonumber\\&\hspace{1cm}=\frac{1}{\pi} D \frac{\beta |J_1-J_2|}{[(E_{\mu_1}^{J_1\pi_1}-E_{\mu_2}^{J_2\pi_2})^2+
\beta^2(J_1-J_2)^2]},
\end{alignat}
where overlines denote ensemble averaging.
Here $J_1\neq J_2$ are the compound nucleus total spin values, $\pi_1,\pi_2$ are parity values,
$E_\mu^{J\pi}$ are resonance energies with $\mu$ being running indices, and $D$ is average level spacing of the compound nucleus.
The $a(b)$ indices specify the orbital momenta $l_{a_{1,2}}(l_{b_{1,2}})$ , the channel spins   $j_{a_{1,2}}(j_{b_{1,2}})$, 
and the microstates $\bar a (\bar b )$ of the target nucleus and
residual nucleus, respectively. Accordingly, $\bar a_1=\bar a_2 $ denote the ground state of the target, and
$\bar b_1=\bar b_2 $ specify the microstates of the residual nucleus.
The above correlation between the partial width amplitudes leads to a correlation between
fluctuating compound nucleus $S$-matrix elements carrying different total spin values:
\begin{equation}
\langle S_{a_1b_1}^{J_1\pi_1}(E)^\ast S_{a_2b_2}^{J_2\pi_2}(E)\rangle=\frac{[\langle|S_{a_1b_1}^{J_1\pi_1}(E)|^2\rangle \langle|S_{a_2b_2}^{J_2\pi_2}(E)|^2\rangle]^{1/2}}{
1+|J_1-J_2|\beta / \Gamma^\uparrow }.
\end{equation}
Here $S_{ab}^{J}(E)$ are compound nucleus $S$-matrix elements
with total spin $J$ and the brackets $\langle...\rangle$ denote the energy $E$ averaging. For finite values of 
$\beta / \Gamma^\uparrow$, nonvanishing of the spin off-diagonal correlations in Eq. (2) reflects
nonvanishing of the interference between resonance levels with different total spins upon the energy
averaging.

For the correlation between $S$-matrix elements carrying the same total spin and parity values and the same
microstates $\bar a_1=\bar a_2$ and   $\bar b_1=\bar b_2$
but different
orbital momenta and/or channel spins we have \cite{kun:94,kun:97b}
\begin{equation}
\langle S_{a_1b_1}^{J\pi}(E)^\ast S_{a_2b_2}^{J\pi}(E)\rangle=[|\langle S_{a_1b_1}^{J\pi}(E)|^2\rangle \langle|S_{a_2b_2}^{J\pi}(E)|^2\rangle]^{1/2}.
\end{equation}
The above equation results from a strong correlation between partial width amplitudes $\gamma_\mu^{J\pi a_1(b_1)}$ and
 $\gamma_\mu^{J\pi a_2(b_2)}$ with $\bar a_1=\bar a_2$ and   $\bar b_1=\bar b_2$ but
$l_{a_1}\neq l_{a_2}$, $l_{b_1}\neq l_{b_2}$, $j_{a_1}\neq j_{a_2}$, $j_{b_1}\neq j_{b_2}$. Such a correlation
is referred to \cite{kun:94,kun:97b} as the continuum correlation. Note that such a strong correlation between
reduced width amplitudes corresponding to the same total spin and parity values but different orbital momenta
was experimentally revealed for a number of compound nuclei in the regime of isolated resonances \cite{Mitchell85}.

For $\beta \gg \Gamma^\uparrow $, the spin off-diagonal correlations in Eq. (2) result in the angular distributions
symmetric around 90$^\circ$ c.m. recovering a conventional picture of compound nucleus. 
However, if  $\beta\leq \Gamma^\uparrow $, {\sl i.e} the phase relaxation time $\tau_{\rm ph}$ 
is comparable or longer than the average life-time of the compound nucleus, this allows us to describe a strong asymmetry
of the angular distributions around 90$^\circ$ c.m. of the evaporating yield. 

For the treatment presented here we follow Ref. \cite{kun:97a}. We neglect the intrinsic spins of the scattering partners in the entrance channel and proton spin in the exit channel. 
Since evaporated protons carrying orbital momenta $l > 1$ are significantly sub-barrier due to both the centrifugal and Coulomb barriers, we take $T_{l > 1}=0$, where $T_l$ are the transmission coefficients for the inverse process of capture of the proton by the residual nucleus. Then the double-differential cross section in the evaporation domain of the spectra has the form \cite{kun:97a}
\begin{equation} \frac{d^2\sigma}{d\Omega d\varepsilon} =\frac{1}{4\pi}\sigma(\varepsilon)
\sum_{L=0}^2 A_L P_L(\cos\theta).
\end{equation}
Here, $\sigma(\varepsilon)$ is the angle-integrated  cross section for the evaporation of a proton with the energy $\varepsilon $, and $P_L(\cos\theta)$ denote the Legendre polynomials of order $L$. The coefficients of the angular decomposition are given by 
\begin{alignat}{1} 
A_{L=0}&=1\nonumber\\ 
A_{L=1}&=\frac{5\sqrt{\frac{T_1}{T_0}}}{\left(1+3\frac{T_1}{T_0}\right)\left(1+\frac{\beta}{\Gamma^\uparrow} \right)}, \nonumber\\ 
A_{L=2}&=\frac{3\frac{T_1}{T_0}}{\left(1+3\frac{T_1}{T_0}\right)\left(1+2\frac{\beta}{\Gamma^\uparrow} \right)}. 
\end{alignat}
One can see that
if this phase memory time $\tau_{\rm ph}=\hbar/\beta$ is about or longer than the average life-time $\tau^\uparrow=\hbar/\Gamma^\uparrow$ of the compound nucleus the evaporation yield is emitted asymmetrically about $90^\circ$ c.m., {\it i.e.} the memory about the direction of the initial beam remains. However, if the phase memory time is much shorter than the average life-time of the compound nucleus then the spin off-diagonal correlations vanish, the memory on the direction of the initial beam is lost, and an isotropic angular distribution around $90^\circ$ c.m. is obtained. 

We now turn to the experimental data available from Ref.~\cite{gugelot} in order to estimate bounds for the possible values of $\beta$. Since in Ref. \cite{gugelot} only relative yields for two scattering angles are reported we focus on the analysis of the shape of the angular distributions $I(\theta )=K d^2\sigma /d\Omega d\varepsilon $ without paying attention to the angle and the energy independent prefactor $K$.

\begin{figure}
\includegraphics[width=8cm]{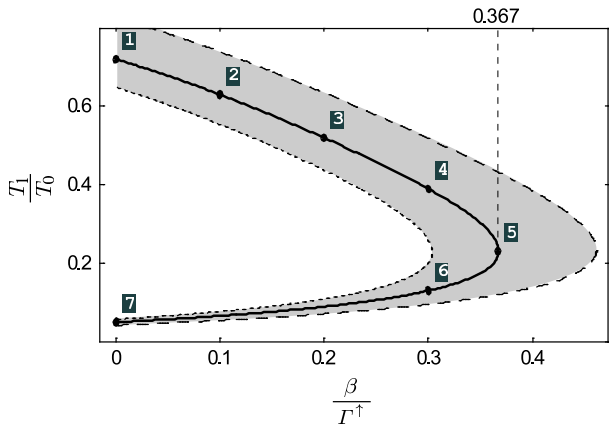}
\caption{\label{fig:impcurves}
Dependence of $T_1/T_0$ on $\beta/\Gamma^\uparrow$ obtained from Eq. (6) for ${\mathcal R}_I=6$ (solid line), $5$ (dashed line), and $7$ (dotted line). The shadowed area shows the range of possible $T_1/T_0$ and $\beta/\Gamma^\uparrow$ which are in accordance with the experimental data of Ref. \cite{gugelot}. The dots marked with numbers on the solid line correspond to {\tt\bf 1}: $\left(\frac{\beta}{\Gamma^\uparrow},\,\frac{T_1}{T_0}\right)$=(0, 0.72); {\tt\bf 2}: (0.1, 0.63); {\tt\bf 3}: (0.2, 0.52);  {\tt\bf 4}: (0.3, 0.39);  {\tt\bf 5}: (0.367, 0.231);  {\tt\bf 6}: (0.3, 0.13);  {\tt\bf 7}: (0, 0.05). }
\end{figure}

For $\varepsilon=7$ MeV, we find from the experimental data  a ratio $I(\theta =60^\circ )/I(\theta =150^\circ )\approx 6$.  Thus, we consider the relation 
\begin{equation} {\mathcal R}_I\equiv\frac{I(\theta =60^\circ )}{I(\theta =150^\circ)}=6, \label{eq:ratio} 
\end{equation} 
where $I(\theta )$ is given by Eqs. (4) and (5) and depends on the two parameters $\beta/\Gamma^\uparrow $ and $T_1/T_0$. 
Therefore, Eq.~(\ref{eq:ratio}) implicitly defines a curve in the parameter space of $\beta/\Gamma^\uparrow $ and $T_1/T_0$, which is plotted in Fig.~\ref{fig:impcurves}. We can either solve Eq.~(\ref{eq:ratio}) for $\beta/\Gamma^\uparrow$ as a function of $T_1/T_0$ or vice versa. In any case we have an underdetermined equation demonstrating that measurements of the evaporation yields for two angles only does not allow an unambiguous determination of the phase relaxation time. Indeed, in  Fig.~\ref{fig:impcurves}
any of the seven dots as well as any point on the solid line corresponds to  ${\mathcal R}_I=6$. Yet, moving along the solid line,  
$\beta/\Gamma^\uparrow$ changes from 0  to 0.367. Since the value of  $T_1/T_0$ can be obtained from model calculations 
this clearly demonstrates that measurement of the evaporation yields for two
angles only 
 permits to accurately determine only the upper experimental limit, $\beta/\Gamma^\uparrow =0.367 $. Estimating the total decay width from the systematics in Fig. 7 of Ref.~\cite{ericson} we obtain $\Gamma^\uparrow = 0.02$ keV and, therefore, $\beta \leq 7$ eV. Taking into account that $\Gamma_{spr}\simeq 1.5$ MeV \cite{flores} we observe that the phase relaxation time is at least five orders of magnitude longer than the energy relaxation time. Yet, it is still about fifteen orders of magnitude shorter than the Heisenberg time, $\hbar /D$, where the average
level spacing of the compound nucleus, $D\sim 10^{-20}$ MeV, has been estimated using the Fermi-gas model \cite{nucphys}.

Due to the uncertainties of the data reported in Ref. \cite{gugelot}, we also included curves of $T_1/T_0$ and $\beta/\Gamma^\uparrow $ in Fig.~\ref{fig:impcurves} resulting from Eq. (\ref{eq:ratio}) but with its right hand side being $5$ and $7$. 

For a more accurate determination of the $\beta$, rather than just an estimation of its upper limit, one has to measure the evaporation yields for more than two angles. Therefore, we analyze the sensitivity of the shapes of the angular distributions for different sets of $\beta/\Gamma^\uparrow $ and $T_1/T_0$ belonging to the manifold obtained from the solution of Eq. (6).

\begin{figure}
\includegraphics[width=8cm]{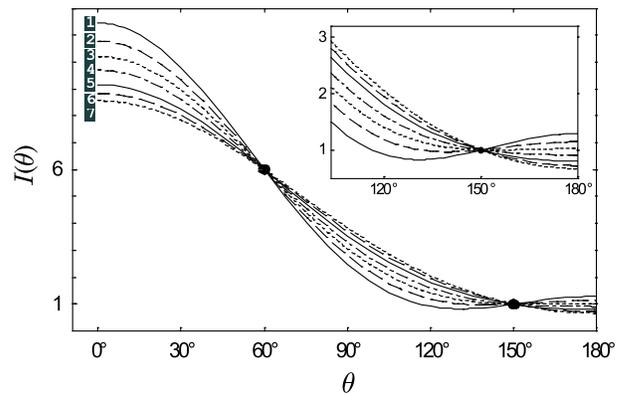}
\caption{\label{fig:angdist}
Angular distributions for the different sets of $\beta/\Gamma^\uparrow$  and $T_1/T_0$ as specified in Fig. 1. In the inset a magnification is shown for the angular range we used for our analysis (see text).
}

\end{figure}

In Fig. 2 we present the angular distributions for the seven sets of $\beta/\Gamma^\uparrow $ and $T_1/T_0$ marked with dots and corresponding numbers in Fig. 1. The angular distributions are normalized in such a way that $I(\theta =60^\circ )=6$ and $I(\theta =150^\circ )=1$ for each of these curves.
One does observe that the angular distributions change appreciably with the change of $\beta/\Gamma^\uparrow $ and $T_1/T_0$ values. To quantify this sensitivity we determine $\theta_{opt}$ for which the ratio $I(\theta_{opt})/I(\theta = 170^\circ )$ is most sensitive to different  $\beta/\Gamma^\uparrow $ and $T_1/T_0$ values. We find that, for ${\mathcal R}_I=6,5,7$, $\theta_{opt}=118^\circ , 116.3^\circ , 119.4^\circ $, respectively.

\begin{figure}
\vspace{5mm}
\includegraphics[width=8cm]{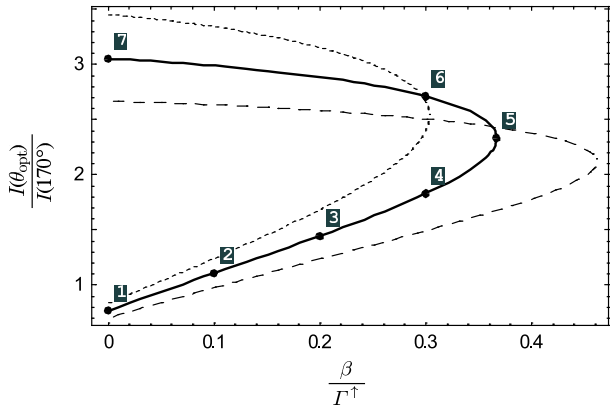}
\caption{The dependencies of $I(\theta_{opt})/I(\theta = 170^\circ )$ on $\beta/\Gamma^\uparrow$. The solid line is obtained for ${\mathcal R}_I=6$ (see Eq. (\ref{eq:ratio})) the dashed line for ${\mathcal R}_I=5$, and dotted line ${\mathcal R}_I=7$. Dots marked with numbers on the the solid line correspond to the $\beta/\Gamma^\uparrow$ and $T_1/T_0$ values specified in Fig. 1. \label{fig:sens}} 
\end{figure}

The dependence of $I(\theta_{opt})/I(\theta = 170^\circ)$ on $\beta/\Gamma^\uparrow$ is presented in Fig. 3 with the solid line corresponding to the ratio ${\mathcal R}_I=6$.  
We find that for values of $I(\theta_{opt})/I(\theta = 170^\circ )\leq 2.7-2.8$ the ratio $I(\theta_{opt})/I(\theta = 170^\circ )$ is rather sensitive to the $\beta/\Gamma^\uparrow$, as can be seen in the lower part of Fig.~\ref{fig:sens}. If the experimental value lies in this range, an accuracy of $I(\theta_{opt})/I(\theta = 170^\circ )$ of about 5$\%$ allows a determination of $\beta/\Gamma^\uparrow $ with a minimal uncertainty of about $10\%$. For a too low experimental value of $I(\theta_{opt})/I(\theta = 170^\circ )$, say less than $0.84$, one can determine only the upper limit of $\beta/\Gamma^\uparrow\leq 0.02$.

On the other hand, in the upper part of the curve, for $I(\theta_{opt})/I(\theta = 170^\circ )\geq 2.7-2.8$, the ratio $I(\theta_{opt})/I(\theta = 170^\circ )$ shows a rather weak dependence on $\beta/\Gamma^\uparrow $. 
In particular, for an assumed accuracy of the data of about $5\%$, the value of $\beta/\Gamma^\uparrow $ can be determined with an uncertainty of only $\simeq 50-100\%$. If the experimental value of $I(\theta_{opt})/I(\theta = 170^\circ )$ is larger than $2.85$ one can determine only the upper limit, $\beta/\Gamma^\uparrow \leq 0.2$.

Similar conclusions about the sensitivity of $I(\theta_{opt})/I(\theta = 170^\circ )$ to $\beta/\Gamma^\uparrow$ can be drawn from Fig. 3 for ${\mathcal R}_I=5$ and $7$. 
According to Fig.~\ref{fig:angdist} an improvement of the accuracy might be found if the data for forward scattering ($\theta < 30^\circ$) would also be available. 

It should be noted that a manifestation of a formation of the thermalized non-equilibrated matter has been
also identified from a strong asymmetry around 90$^\circ$ c.m. of evaporating protons in the Bi($\gamma$,p)
photonuclear reaction, see \cite{sigma} and references therein. Other examples of a strong forward peaking of evaporating protons are
found, {\sl e.g.}, in Bi(p,p') and Bi(n,p) processes with the 62 MeV energy of initial beam, see \cite{flores} and references therein. 
Even though the later examples do demonstrate a formation of the thermalized non-equilibrated matter in compound processes,
a determination of the phase relaxation time from these data is not unambigous since, for such high energy of the initial
beam, there is a high probablility for a second and third chance proton evaporation.  

The possibility that in highly excited many-body systems
the phase relaxation can be much longer than the energy relaxation may have significant implications for quantum computing
\cite{flores,sigma} as well as, {\sl e.g.}, time-delayed ``statistical'' ionization of many-electron quantum dots and 
atomic clusters (see, {\sl e.g.}, \cite{campb} and references therein).
A possible presence of the effect of anomalously slow phase relaxation \cite{Luis} in chemical reactions (see  \cite{Luis}
and references therein) would require a 
modification
of the statistical theories - phase space and transition state theories (see, {\sl e.g.}, \cite{Levine} and references therein). 
Yet, the nuclear data indicating
an existence of anomalously slow phase relaxation, which is much slower than the energy relaxation, are largely unrecognized
by nuclear physicists and 
unknown outside the nuclear physics community. In many fields, including statistical physics, 
the notion ``thermalization'' or ``energy equilibration'' is considered to be equivalent to the 
notion ``statistical equilibrium''. This note is a step towards changing this undesirable situation.

In conclusion we have proposed a general method to  estimate the accuracy of the determination of the phase relaxation time from data on forward peaking of evaporating protons from compound nucleus.  
The angular range and accuracy of the data needed for a reliable determination of the phase relaxation time have been evaluated. The general method has been applied to the analysis of inelastic scattering of 18 MeV protons from Pt for which previously measured double differential cross sections for two angles in the evaporating domain of the spectra show a strong forward peaking. 
We found that a new measurement of the angular distributions of evaporating protons in the Pt(p,p') inelastic scattering for a wider angular range should permit an accurate determination of the phase relaxation time. 
The experiment is also highly desirable for an accurate test of the formation of thermalized non-equilibrated matter in compound processes.
Our analysis for the improvement of the experimental determination of the phase relaxation time from nuclear evaporation data can readily be applied to any low energy nucleon-nucleus scattering showing forward peaking in the evaporation domain of the spectra.

\begin{acknowledgments}
The work on the revised version of this paper has been carried out during the the workshop ``Phase relaxation versus energy relaxation in quantum many-body systems'' at the International Center for Science (CIC), Cuernavaca, Mexico, October--November 2005. We are grateful to Bertrand Georgeot, Garry Mitchell, Thomas Seligman, Hans Weidenm{\"u}ller and Wang Qi for encouraging discussions and for pointing out that the proposed experiment is highly desirable for an accurate test of the formation of thermalized non-equilibrated matter in quantum many-body systems. These discussions and suggestions have stimulated the extension of the original version of the paper.
Marc Bienert thanks the Alexander--von--Humboldt foundation for the support of a Feodor--Lynen fellowship.
\end{acknowledgments}

\end{document}